\documentclass[runningheads,a4paper]{llncs}

\usepackage{amssymb}
\usepackage{graphicx}
\usepackage{amssymb}

\usepackage{units}

\usepackage{url}
\urldef{\mailsa}\path|{nick.fyson, tijl.debie,|
\urldef{\mailsb}\path|nello.cristianini}@bris.ac.uk|
\newcommand{\keywords}[1]{\par\addvspace\baselineskip
\noindent\keywordname\enspace\ignorespaces#1}

\usepackage[labelfont={sf,bf}, textfont=sf]{subfig}

\usepackage{algorithmic}
\usepackage{algorithm}

\usepackage{booktabs}

\begin{document}

\mainmatter  

\title{Reconstruction of Causal Networks \\ by Set Covering}

\titlerunning{Reconstruction of Causal Networks by Set Covering}

\author{Nick Fyson\inst{1,2}%
\and Tijl De Bie\inst{1}
\and Nello Cristianini\inst{1}
}
\authorrunning{Reconstruction of Causal Networks by Set Covering}

\institute{
Intelligent Systems Laboratory, Bristol University\\
Merchant Venturers Building, Woodland Road, Bristol, BS8 1UB, UK\\
\and
Bristol Centre for Complexity Sciences, Bristol University, \\
Queen's Building, University Walk, Bristol, BS8 1TR, UK \\
\url{http://patterns.enm.bris.ac.uk}
}

%
%

\toctitle{Reconstruction of Causal Networks by Set Covering}
\tocauthor{Nick Fyson, Tijl De Bie, Nello Cristianini}
\maketitle

\begin{abstract}

We present a method for the reconstruction of networks, based on the order of nodes visited by a stochastic branching process. Our algorithm reconstructs a network of minimal size that ensures consistency with the data. Crucially, we show that global consistency with the data can be achieved through purely local considerations, inferring the neighbourhood of each node in turn. The optimisation problem solved for each individual node can be reduced to a Set Covering Problem, which is known to be NP-hard but can be approximated well in practice. We then extend our approach to account for noisy data, based on the Minimum Description Length principle. We demonstrate our algorithms on synthetic data, generated by an SIR-like epidemiological model.

\end{abstract}

\section{Introduction}

There has been increasing interest over recent years in the problem of reconstructing complex networks from the streams of dynamic data they produce. Such problems can be found in a highly diverse range of fields, whether determining Gene Regulatory Networks (GRNs) from expression measurements \cite{Sprinzak:2005p5798}, or the connectivity of neuronal systems from spike train data \cite{Brown:2004p6290}. While data in the the field of GRNs is generally continuous in nature, spike train data is inherently discrete. Other fields include epidemiology, chemical engineering and manufacturing \cite{Unnikrishnan:2006p3140}, but all share the similar challenge of extracting the causal structure of a complex dynamical system from streams of temporal data.

We here address the challenge of reconstructing networks from data corresponding to stochastic branching processes, occurring on directed networks and where a discrete `infection' is propagated from node to node. The clearest analogy lies in the field of epidemiology, where instances of infection begin at particular nodes, before propagating stochastically along edges until the infection dies out. Another source of such data could be blogs, where initial report of a story is made on a particular site, before being picked up by other blogs and `cascading' through the blogosphere \cite{Leskovec:2007p5462}. Analysis of such data could permit the reconstruction of a network of readership. Most generally, we could consider data corresponding to `memes', fundamental units of cultural information which propagate through all systems of communication, notably the news media system.

The main contributions of this paper are:

\begin{enumerate}
\item \textbf{A novel approach to the reconstruction of networks from data corresponding to stochastic branching processes.} We clearly define the form of the data and optimisation problem, before reducing it to the well-known Set Covering problem.
\item \textbf{A modification extending our approach for use on noisy data.} We use the concept of Minimum Description Length (MDL) to define a criterion for halting greedy set covering, allowing us to reconstruct networks from data containing lost entries.
\end{enumerate}

The paper is organised as follows. In Sec.~\ref{sec:network_reconstruction} we fully define the nature of data to be used, the problem we address, and the optimisation problem to be tackled. Section \ref{sec:theoretical_analysis} presents a theoretical analysis of our basic algorithm, before Sec.~\ref{sec:noisy_data} introduces an extension to address noisy data, based on the concept of Minimum Description Length (MDL). Section \ref{sec:empirical_evaluation} presents an empirical analysis, before we outline our conclusions in Sec.~\ref{sec:conclusions}.

\section{Network Reconstruction} \label{sec:network_reconstruction}

\subsection{Concepts and Notation} \label{sec:network_reconstruction:concept_notation}

A directed network $G$ is defined by a set of nodes $V$ and a set of oriented edges $E \subseteq V \times V$ between these nodes, and we denote it as $G = (V,E)$. In this paper we consider two networks over the same set of nodes. $G_T = (V,E_T)$ is the true underlying network, while the reconstructed network we infer from data is denoted by $G_R = (V,E_R)$. We assume a dynamic branching process occurs on the network $G_T$, in which the transfer of `markers' occurs. Markers originate at a particular node in the network, and then propagate stochastically from node to adjacent node, `traversing' along only those edges that exist in the set $E_T$.

With analogy to terminology in the field of epidemiology, we refer to the process of a node becoming a carrier of a marker as \textit{infection}. Similarly, all nodes that have undergone infection at any point in the past and remain in a state where they are \textit{infectious} are known as \textit {infected}. Finally, any node that underwent infection at any point from a particular marker is referred to as a \textit{carrier}.

Each marker that is propagated through the network generates a `marker trace', $M^i$. The set of all marker traces is denoted by $\:\:{\cal M} = \{ M^i \}$, and throughout the paper we use superscripts to index between markers. The marker trace is represented by an ordered set of the nodes that carried that marker, in the order in which they became infected. We will use subscripts to refer to individual nodes in a marker trace. We formally define the notion of a marker trace as follows.

\begin{definition}[Marker Trace, $M^i$] \label{def:marker_trace}  A Marker Trace $M^i$ is an ordered set of $n_i$ distinct nodes $w^i_j \in V$, and we denote it as:
\begin{eqnarray*}
M^i &=& (w^i_1,w^i_2,\ldots,w^i_{n_i})
\end{eqnarray*}
\end{definition}

Each marker trace defines a total order over the reporting nodes, and we use the notation $v_i <_{M^i} v_j$ to state that the node $v_i$ appears before node $v_j$ in the marker trace $M^i$.

For clarity in future definitions we also formally define a path from one node to another within a network.

\begin{definition}[Path in a network $G = (V,E)$] \label{def:path} A sequence $U=u_1,\ldots,u_k$ of nodes $u_i \in V$ is a path in $G=(V,E) \:$ if   $\: \forall \:\:1\le i < k \: , (u_i,u_{i+1}) \in E$
\end{definition}

\subsection{Problem Formulation and Global Consistency} \label{sec:network_reconstruction:prob_formulation}

\begin{problem}[Informal Description] \label{prb:informal_def} Given a set $\cal M$ of Marker Traces construct a network $G_R$, approximating the true network $G_T$ that generated $\cal M$.
\end{problem}

Intuitively, it makes sense to choose $G_R$ such that it is capable of generating $\cal M$ itself as well. Given our assumptions on the mechanism of data generation, this requires that for each marker a path exists from the originator to all other carrier nodes, passing only through nodes that have been previously infected. We will refer to this as `global consistency' and formalise the intuitive notion as follows.

\begin{definition}[Globally Consistent, GC] \label{dfn:global_consistency}
\begin{eqnarray*}
G_R \mbox{ is GC with } M^i    \iff  \:\:    \forall w^i_j \mbox{ with } j>1    \:\:\:\:\:\:     \exists\:\: \mbox{ a path } \:\:\: w^i_1, \ldots, w^i_j  \:\mbox{ in }\: G_R
\end{eqnarray*}
\end{definition}

Besides being intuitively satisfying, in Sec.~\ref{sec:theoretical_analysis} we will prove that ensuring global consistency also ensures that a large number of edges from $G_T$ is guaranteed to be reconstructed in $G_R$. Trivially, it is clear that a completely connected network is consistent with all possible data, and hence we aim to reconstruct a consistent set $E_R$ of minimal size.

Combining the above allows us to formalise our goal in terms of an optimisation problem.

\begin{problem}[Formulation in terms of Global Consistency] \label{prb:global_consistency}
\begin{eqnarray*}
\mathrm{argmin}_{E_R} |E_R|
\end{eqnarray*}
subject to
\begin{eqnarray*}
&& \forall \:\:\:\: M^i \in {\cal M} \:\:\:\: \:\:\:G_R =(V,E_R) \textrm{ is GC with }  M^i
\end{eqnarray*}
\end{problem}

\subsection{Local Consistency}

For a reconstruction to make intuitive sense we require global consistency between network and data, but this involves consideration of paths and is impractical. Below, we demonstrate the equivalence of global consistency with `local consistency', an alternative that allows us to consider the immediate neighbourhood of each node in turn.

Local consistency requires that for each node reporting a particular marker, the node must have at least one incoming edge from a node that has reported the marker at an earlier time. This concept is formalised as follows.

\begin{definition}[Locally Consistent, LC] \label{dfn:local_consistency}
\begin{eqnarray*}
G_R \textrm{ is LC with } M^i \iff  \:\: \forall w^i_j \mbox{ with } j>1 \:\:\:\:\:\:\: \exists \:\: w^i_k \mbox{ with } k < j : (w^i_k,w^i_j) \in E_R
\end{eqnarray*}
\end{definition}

\begin{theorem} [LC $\iff$ GC] \label{clm:lc_iff_gc}
Demonstrating local consistency between $G_R$ and $M^i$ is necessary and sufficient to ensure global consistency.
\end{theorem}
\begin{proof}
We define an approach to constructing a network that ensures every node has an incoming edge from a node that reported at an earlier time (local consistency), and demonstrate that this necessarily ensures that a path exists from the originator to every other node (global consistency).

For the case $k=1$, we have only the originator node, hence trivially there is a path from originator to all other nodes.
For the case $k=2$, we add a node with an incoming edge from the only other node. Again trivially, there is a path from the originator to every other node. For the case $k=n+1$ we take the network for $k=n$, and add a node with an incoming edge from one of the existing nodes. If there is a path from originator to all nodes in the $k=n$ network, there will be a path from originator to the new node in the case $k=n+1$. Hence if the claim is true for $k=n$ then it is also true for $k=n+1$.

Therefore, by induction,  LC $\iff$ GC. $\qed$
\end{proof}

This allows us to formulate an alternative but equivalent optimisation problem, using using the concept of local consistency.

\begin{problem}[Formulation in terms of Local Consistency] \label{prb:local_consistency}Á
\begin{eqnarray*}
\mathrm{argmin}_{E_R} |E_R|
\end{eqnarray*}
subject to
\begin{eqnarray*}
&& \forall \:\:\:\: M^i \in {\cal M} \:\:\:\: \:\:\:G_R =(V,E_R) \textrm{ is LC with }  M^i
\end{eqnarray*}
\end{problem}

Crucially, to establish local consistency, one need only consider the immediate neighbourhood of each node in turn. Hence we can break this optimisation problem into $N$ subproblems, where $N$ is the total number of nodes in the network. In each of these subproblems, we establish the minimal set of incoming edges required to explain all the markers reported by the particular node. From now on, unless otherwise specified, we describe approaches as applied to discovering the parents of a particular node, which would then be applied to each node in turn.

\subsection{Formulation in terms of Set Covering}  \label{sec:network_reconstruction:form_set_covering}

Using the concept of local consistency we are able to treat the reconstruction on a node-by-node basis, and we denote the node under consideration as $v$.  As specified by local consistency, in considering the incoming edges for a particular node we must include at least one edge from a node that has reported each marker at an earlier time. Each edge therefore `explains' the presence of a subset of the reported markers, and if the set of all incoming edges together explains all the reported markers, we ensure local consistency. This problem of `explaining' marker reports may be neatly expressed as a Set Covering Problem.

Before showing how it relates to our reconstruction problem, we formally state the Set Covering optimisation problem: Given a universe $\cal A$ and a family $\cal B$ of subsets of $\cal A$, the task is to find the smallest subfamily ${\cal C} \subseteq {\cal B}$ such that $\bigcup {\cal C} = {\cal A}$. This subfamily $\cal C$ is then the `minimal cover' of $\cal A$. Given this formal framework, we now define how these sets relate to our reconstruction problem.

\begin{definition}[Universe, ${\cal A}^v$] \label{dfn:set_to_be_covered_A} The universe set of all elements is defined as the set of all markers that have been reported by the node $v$:
\begin{eqnarray*}
{\cal A}^v = \{ i  : v \in M^i \}
\end{eqnarray*}
\end{definition}

The node $v$ can have an incoming edge from any other node, and hence the space of potential incoming edges is ${\cal F}^v = \: (V  / v ) \: \times \: v$. As stated above, each potential incoming edge will `explain' a subset of the markers reported by $v$, and therefore every edge $f^v_j \in {\cal F}^v$ corresponds to one element $B^v_j$ in the family of subsets ${\cal B}^v$.

\begin{definition}[Family of subsets, ${\cal B}^v = \{ B^v_j \}$] \label{dfn:covering_sets} Each subset $B^v_j$ is defined by a potential incoming edge $(v_j,v) = f^v_j \in {\cal F}^v$, where $i$ is in $B^v_j$ if and only if $v_j$ appears earlier than $v$ in the marker trace $M^i$:
\begin{eqnarray*}
B^v_j &=& \{ i \:\:    :    v_j <_{M^i} v \}
\end{eqnarray*}
\end{definition}

The set covering problem then requires us to find a subfamily ${\cal C}^v \subseteq {\cal B}^v$ such that $\bigcup {\cal C}^v = {\cal A}^v$, and this subfamily ${\cal C}^v$ directly corresponds to a set of incoming edges for the node $v$.



\begin{definition}[Reconstructed Incoming Edges, $E^v_R$] \label{dfn:reconstructed_incoming_edges} The set of reconstructed edges, $E^v_R$, consists of the set of all elements in $\cal F$ that correspond to elements of $\cal C$:
\begin{eqnarray*}
E^v_R &=& \{ f^v_j \in {\cal F}^v : B^v_j \in {\cal C}^v  \}
\end{eqnarray*}
\end{definition}

This then allows us to make a final definition of our optimisation problem, this time in terms of the Set Covering Problem. The following problem is defined for each node $v \in V$.

\begin{problem}[Formulation in terms of Set Covering] \label{prb:set_covering}
\begin{eqnarray*}
\mathrm{argmin}_{E^v_R} |E^v_R|
\end{eqnarray*}
subject to
\begin{eqnarray*}
&& A^v = \bigcup {\cal C}^v \:\:\:\: \mbox{ where } \:\: {\cal C}^v = \{ B^v_j : f^v_j \in E^v_R \} \:\: \mbox{ and } \:\: E^v_R \subseteq {\cal F}^v
\end{eqnarray*}
\end{problem}

Finally, repeating this optimisation for all nodes in the network, we get $E_R = \bigcup_v E^v_R$, allowing us to reconstruct the entire network through only local considerations.

\subsection{Greedy Approximation to Set Covering} \label{sec:greed_approx}

The Set Covering Problem is known to be NP-hard, but in practice is easy to approximate well using a greedy approach (see Sec.~\ref{sec:theoretical_analysis}). The greedy algorithm is well documented for set covering \cite{Chvatal:1979p5458}, but below we briefly outline the approach.

We wish to cover the set ${\cal A}$ by selecting from the family of subsets ${\cal B}$. We first select the subset $B_j \in {\cal B}$ that covers the greatest number of elements in ${\cal A}$, ie. such as to maximise $|B_j|$. The corresponding edge $f_j$ is then added to the set of reconstructed edges $E^v_R$. A subset of ${\cal A}$ has now been covered, and hence these elements are removed both from ${\cal A}$ and all subsets in the family $\cal B$. This process is repeated until ${\cal A} = \emptyset$.


\section{Theoretical Analysis} \label{sec:theoretical_analysis}

We have formalised our problem using the intuitive notion of global consistency of a network with a set of Marker Traces. Here, we will show that this strategy ensures that the reconstructed network $G_R$ is close to the true network $G_T$ in a well-defined sense. In particular, we will consider the number of edges in $E_R$ also in $E_T$, referred to as the True Positives (TP), as well as the number of edges in $E_R$ not in $E_T$, referred to as the False Positives (FP). The number of true and false positives gives an indication both of how well the approach finds edges that really exist, and how likely it is to incorrectly identify edges as being part of the network.

\subsection{Lower bound on True Positives}

The number of true positives found by the reconstruction approach is simply the number of edges found in both the true and reconstructed networks, given by

\begin{equation}
\mbox{TP} = | E_R \cap E_T |
\end{equation}

The nature of the set covering algorithm allows us to set a lower limit on TP, given a particular ${\cal M}$. In achieving a complete coverage we can be certain of including all those edges that were traversed first in the propagation of any marker. In other words, all pairs of nodes that appear first and second in any marker trace are guaranteed to represent a true edge, and will also inevitably be included in the covering. There is only a single subset covering this particular marker report, an hence it must be included in the final reconstruction. We therefore only need count the number of such edges to determine the least number of true edges we will identify, and thus a lower bound $\mbox{TP}_-$  on  $\mbox{TP}$ 

In assessing performance it is useful also to define the True Positive Rate (TPR), which is the fraction of true positives successfully recovered, given by

\begin{equation}
\mbox{TPR} = \frac{| E_R \cap E_T |}{|E_T|}
\end{equation}

Trivially, if $|E_T|$ is known, we can use $\mbox{TP}_-$ to directly obtain a lower bound $\mbox{TPR}_-$ on $\mbox{TPR}$.

\subsection{Upper bound on False Positives}

While we may correctly include all genuine edges, it is also important to successfully exclude all false edges from our reconstruction. This is quantified by the number of False Positives (FP):

\begin{equation}
\mbox{FP}   =  | E_R / E_T |
\end{equation}

We denote the highest possible number of false positives as $\mbox{FP}_{+}$, and in order to specify this bound we need to make the following definitions. The set $E_R$ is obtained from a greedy approximation to set covering, and therefore is not guaranteed to be optimal. We denote the optimal covering as $E_R^*$, which will always be equal or smaller in cardinality than $E_R$. We also know that the true set of edges will always provide a valid covering, and hence provides an upper bound on the size of the optimal covering, giving $|E_R^*| \le |E_T|$. Finally, the heuristic ratio is defined as the upper bound on the size of the obtained set relative to the size of the optimal set, $H \ge |E_R|/|E_R^*|$.

We can now specify the upper bound on false positives as follows:
\begin{eqnarray}
\mbox{FP} \:\:\:\:\:\:\:\:  		   &=&\:\:\:\: \:\: |E_R| - \:\: \mbox{TP} \\
\mbox{FP} \:\:\:\:\:\:\:\:  		   &\le&\:\:\:\: \:\: |E_R| - \:\: \mbox{TP}_- \\
			  					   &\le&\:\:\:\:H . |E_R^*| -  \mbox{TP}_- \\
			  					   &\le&\:\:\:\: H . |E_T| -  \mbox{TP}_- \\
\therefore \:\:\:\:\mbox{FP}_+ \:\:\:\:\:  &=&\:\:\:\: H . |E_T| -  \mbox{TP}_-
\end{eqnarray}

The greedy approximation to set covering is known to be as good as any polynomial-time approximation, and the literature gives us two useful values for $H$. The first bound is related to the maximum size of the subsets from which we construct the covering, $\max_{B_j \in B}|B_j|$, and provides a limit on the quality of the covering related to the logarithm of the size of this set \cite{Chvatal:1979p5458}. This first bound $H_1$ is given by

\begin{equation}
H_1 = 1+\ln \left(  \max_{B_j \in B}|B_j|   \right)
\end{equation}

The second bound approaches the problem from an alternative perspective, considering the maximum number of covering subsets of which any element is a member, $m_0$ \cite{Slavik:1997p5625}. In other words, for each element of the ground set we need to cover, how many subsets can be selected from in order to cover the element in question. In the case of our algorithm this is related to the length of marker traces. The elements of the ground set we need to cover are reports of a marker at a node, and the number of ways of explaining the presence of this marker is equal to the number of nodes that have reported at an earlier time. The maximum membership across all elements in the ground set, $m_0$, is therefore related to the maximum length of marker traces. This second bound is then

\begin{eqnarray}
H_2 &=& m_0 \\
 &=& \left( \max_{M^i \in {\cal M}} | M^i | \right) - 1
\end{eqnarray}

Initially $H_2$ appears to provide a less useful bound, since it is linear as opposed to logarithmic, but the behaviour of the bounds as the number of markers increases is markedly different. While the maximum size of covering set continues to increase with number of markers, the maximum length of marker trace rapidly tends to a fixed value. This limit is a property of the network and marker propagation, but at most is limited by the size of the network, not the dataset. Therefore, as the amount of data used in the reconstruction increases, the tighter bound switches from $H_1$ to $H_2$. Hence, we can define the heuristic ratio as the minimum of these two alternatives, and therefore bound the false positives as shown in equation \ref{eqn:fp_plus}. Again, determining this bound requires knowing the set of markers and $|E_T|$:

\begin{eqnarray} 
\mbox{FP}_{+} &=& |E_T| .  \min (H_1, H_2) -  \mbox{TP}_- \label{eqn:fp_plus}
\end{eqnarray}

The total number of possible false positive is given by $(|V|^2-|V|)- |E_T|$, and hence we can also define an upper bound on the False Positive Rate ($\mbox{FPR}_{+}$):

\begin{eqnarray} 
\mbox{FPR}_{+} &=& \frac{|E_T| .  \min (H_1, H_2) -  \mbox{TP}_-}{(|V|^2-|V|)- |E_T|} \label{eqn:fpr_plus}
\end{eqnarray}

\subsection{Jaccard Distance}

To assess the overall quality of our reconstructions, we require a measure of how well the reconstructed set of edges matches the true set. In comparing two sets over the same elements, it is appropriate to use the Jaccard Distance (JD). For identical sets this has a value zero, and a value of one if the two sets have no elements in common at all. The JD is given by

\begin{eqnarray}
\mbox{JD} = \frac{| E_T \cup E_R | -  | E_T \cap E_R |}{ | E_T \cup E_R |}
\end{eqnarray}

A lower value of Jaccard Distance indicates a closer match between true and reconstructed networks, and hence a bound on worse-case performance is an upper limit on JD. This can be calculated from bounds on the number of true and false positives as follows:

\begin{eqnarray}
\mbox{JD} \:\:\:\: &\leq&  \:\:\:\:  1 - \frac{\mbox{TP}_{-}}{|E_T| + \mbox{FP}_{+}}
\end{eqnarray}

This upper limit on JD is determined given a particular set of marker traces, and constitutes a worst-case scenario for our success in reconstructing the true underlying network.

\section{Reconstruction from Noisy Data} \label{sec:noisy_data}

Our approach and analysis has thus far assumed perfect and noise free data from which to reconstruct networks. In reality this is an unrealistic assumption, and hence we define an adaptation of our approach to accommodate noisy data.

Our basic Set Covering approach assumes that a minimal network consistent with the data will result in a perfect reconstruction (given infinite data and perfect minimal covering set). Every report of a marker is assumed to be due to direct infection from an earlier infected node, and hence we require that the presence of every marker at every node be explained. When noise is present these assumptions do not hold, and missing marker reports may incorrectly suggest the presence of edges that are not really present. This will lead to a large number of false positives, increasing with the quantity of data used in reconstruction.

In executing the greedy approximation to Set Covering, we first select those subsets that cover the greatest number of remaining elements, which in our case corresponds to choosing edges that explain the greatest number of marker reports. While the noise level remains low, therefore, we will first select true edges, since the incorrect edges suggested by the noise will tend to be relatively low in frequency. We can therefore expect that, in general, the noise-induced false positives will be added toward the end of the set covering process. This is demonstrated empirically in Fig.~\ref{fig:jd_mdl_vs_edges_a},  Sec.~\ref{sec:jd_mdl_set_covering}, and motivates the definition of a criterion to halt the covering early.

\subsection{Minimum Description Length} \label{sec:noisy_data:mdl}

In selecting the optimal point to halt the set covering when reconstructing from noisy data, we appeal to the Minimum Description Length (MDL) principle~\cite{Wallace:1968p5644}. This states that in model selection one should prefer models that are able to communicate the data in the lowest number of bits. This is in principle equivalent to considering Maximum Likelihood Estimation \cite{mackay_information_2002}, but our case lends itself particularly well to the use of MDL.

\subsubsection{Marker Trace Coding Scheme}

We choose to describe the network in the most simple way, in which all edges are explicitly assigned 0 or 1, and hence the network is description is of fixed length. As such, our coding scheme contains no inherent preference for sparsity, and the Description Length (DL) is entirely dependent on how efficiently the set of all markers can be expressed.

In order to describe a marker trace we need to specify in order all those nodes that are members of the set $M^i$. A simple ordered list requires $\ln N$ bits of information per node, where $N$ is the number of nodes in the network. This is straightforward, but using the framework of the underlying network may allow us to describe this same information in a compressed form. Instead of simply listing the reporting nodes, we describe the progression of the marker through the network.

When the network is consistent with the data, we are able to describe all markers exactly with the following approach. We first identify the originator node, at a cost of $\ln N$. We then describe each node of the marker by first identifying its parent (from the set of those that have already reported), and then specifying the particular child of this node. The cost of identifying the 2nd report is then $(\ln 1 + \ln d_{p_2})$, where $d_{p_2}$ is the out-degree of the parent. The 3rd report then requires $\ln 2$ bits to specify the parent, since there are two possibilities,  and also $\ln d_{p_3}$ to specify which child. This progresses similarly for all subsequent reports in the trace. By then summing the description lengths of all marker traces we get the cost of describing the set of data completely.

To render this coding scheme useful in practice we need to be able to describe markers that are not consistent with the network, for which we need only make a simple extension, allowing for the coding of `exceptions'. We do this by defining a `supernode' in addition to the standard network, which is the originator of all markers and by definition a parent of every other node. The description of the first report then becomes $(\ln 1 + \ln d_{p_1}) = \ln N$, where the cost of specifying the parent is $\ln 1 = 0$ (since all markers originate at the supernode) and the cost of specifying the child is $\ln N$. For the second report there are now two potential parents, and hence to specify the second reporting node we require $(\ln 2 + \ln d_{p_2})$ bits. If the first reporter is a parent of the second, $d_{p_2}$ will be equal to the out-degree of the first reporter, and otherwise $d_{p_2}=N$, the out-degree of the supernode. Similarly, the cost for the third report is $(\ln 3 + \ln d_{p_3})$ bits, the fourth $(\ln 4 + \ln d_{p_4})$ and so on.

A crucial characteristic of this coding scheme is that, while there is no explicit cost to defining edges, nodes of higher degree are more expensive to use as the parent of a report. Therefore, while it is expensive to describe a report as an exception, there is a trade off between creating a network that does not require any exceptions and the increased cost of describing all of the marker reports. In general, therefore, the network that allows the shortest description of all marker traces will lie at some point between completely disconnected and completely connected.

\subsection{MDL as Stopping Criterion}

To use MDL as a stopping criterion requires a minor change to the set covering reconstruction algorithm, in which the addition of edges is considered globally, rather than simply on a node by node basis. We still perform greedy set covering for each node in turn, but instead of placing selected edges directly into the reconstructed network, we make a note of each edge and the number of additional elements covered when it is selected. After doing this for all nodes we have a list of edges across the whole network, along with their explanatory power within the greedy set covering framework. We then rank them all by the elements covered, and follow this order in adding edges to the network.

While the Jaccard Distance requires knowledge of the true network to calculate, we can calculate the description length using only the data and the current reconstructed network. We can therefore calculate the new DL after each edge is added, and subsequently select the network $E_R$ that gave the lowest total description length.

\section{Empirical Evaluation} \label{sec:empirical_evaluation}

\subsection{Generation of Synthetic Data}

The model for generation of our generalised `markers' is based on an SIR epidemiological model. We simulate each marker separately, dropped at random into the network and subsequently propagated between outlets in a stochastic fashion. The definition of this model then falls into three sections; the network itself, the generation of markers and the model used for noise in the data.

\subsubsection{Network Model}

The definition of the network consists of a non-symmetric binary adjacency matrix, $(i,j)=1$ indicating an edge connecting from node $i$ to node $j$. We use a directed Erd\H{o}s-R\'{e}nyi model, in which each edge exists with probability $p = 2/N$, where $N$ is the total number of nodes. This results in an average of 2 outgoing and 2 incoming edges per node, resulting in a relatively sparse network that is likely to be a single weakly-connected component.

\subsubsection{Marker Generation}
Throughout the simulation, each node can be in one of three states; Susceptible (S), Infected (I) or Recovered (R). All nodes begin in state S, before the marker initially `seeded' at a randomly selected node, set to I. The state of each node in the next time step is determined stochastically from its current state and that of all its parents. The potential transitions of a node and their associated probabilities are shown in table \ref{tab:transition_probs}. We select parameters to generate markers paths of reasonable length and frequency, with $p_I = 0.1$ and $p_R = 0.1$.

\subsubsection{Generation of noisy data} We consider the most basic model of noise for the type of data we are looking at, in which each marker report has a certain probability of being `lost'. The single parameter is $p_{\mbox{loss}}$, giving the likelihood that a carrier node is omitted from the marker.

\begin{table}
\centering
\caption{\textbf{The transition probabilities for nodes, dependent on current state.} $n_I$ is number of incoming edges from infected nodes, $p_I$ is probability that infection will pass along an edge in a time step, and $p_R$ is probability that node will recover from infection in a time step.}
\begin{tabular}{ l l l l l}
\noalign{\smallskip}
\toprule
Susceptible &~~~& Infected &~& Recovered \\
\midrule \noalign{\smallskip}
$P(S) = (1-p_i)^{n_I}$      &~~~~~~& $P(I) = 1- p_R$  &~~~~~~& $P(R) = 1$  \\
$P(I)   = 1 - (1-p_i)^{n_I} $ &~~~& $P(R) = p_R$ &~~~& \\
\noalign{\smallskip} \bottomrule
\end{tabular}
\label{tab:transition_probs}
\end{table}

\subsection{Naive Approaches}
We define two naive algorithms for network reconstruction, to which it will be instructive to compare our Set Covering approach.

\subsubsection{Naive 1} The most immediately obvious explanation for the creation of a marker trace is that each node became infected by that node immediately preceding it in time. Indeed, assuming all network structures are equally likely, and considering a trace in isolation, this would be our best guess. We therefore simply take the union of all edges implied by a literal interpretation of each marker trace. The resultant network is capable of producing the observed data, and hence is consistent. This set of edges is given by

\begin{equation}
E_{N_1} = \bigcup_{M^i \in {\cal M}} \: (w^i_n,w^i_{n+1}) \:\:\: \forall n
\end{equation}

\subsubsection{Naive 2} In the second naive approach to reconstruction, we consider only those marker reports for which only one edge can provide the explanation. In other words, we take only those edges that are guaranteed true positives. This does not make full use of the available information, since it effectively throws away all reports of a marker beyond the second, but ensures no false positives are included. The set of edges given by the second naive method is given by

\begin{equation}
E_{N_2} = \bigcup_{M^i \in {\cal M}} (w^i_1,w^i_2)
\end{equation}

\subsection{Naive vs. Set Covering Approaches - Noise Free Case}

\begin{figure}
  \centering
  \subfloat{
  \includegraphics[width=1\textwidth]{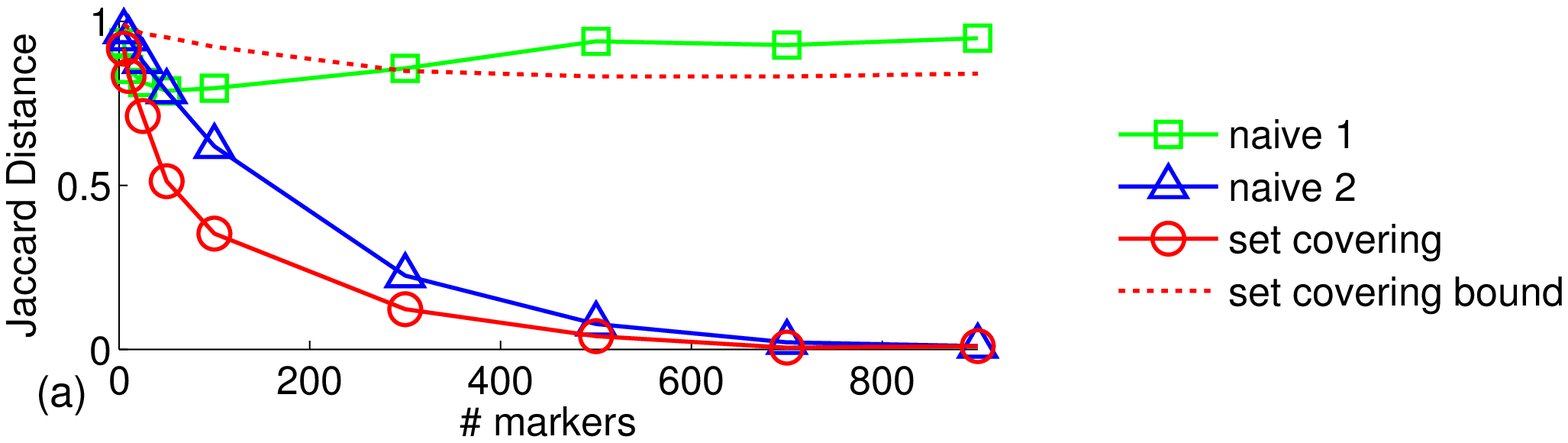} \label{fig:basic_evaluation_a}
  } \\
  \subfloat{
  \includegraphics[width=0.5\textwidth]{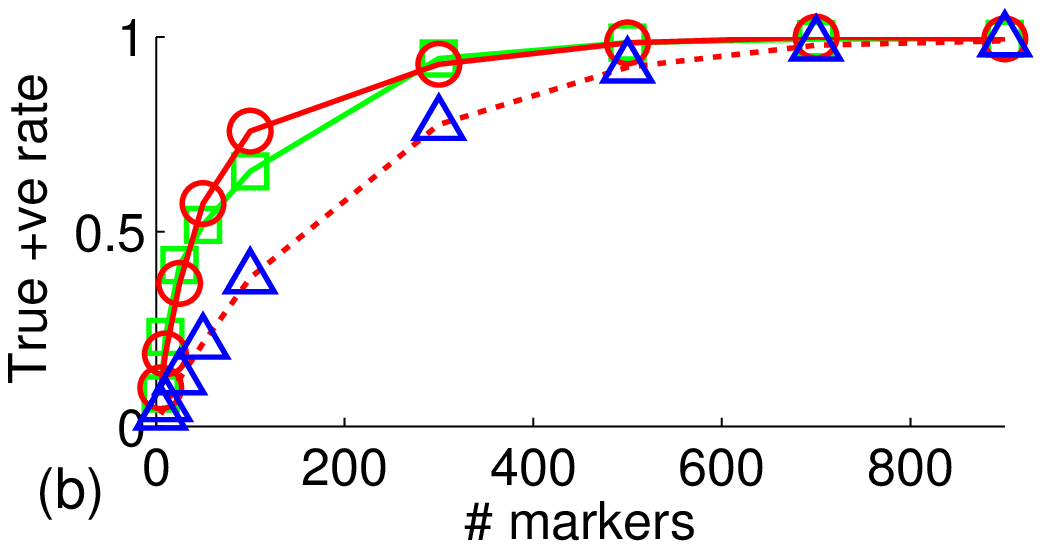} \label{fig:basic_evaluation_b}
  } 
  \subfloat{
  \includegraphics[width= 0.5\textwidth]{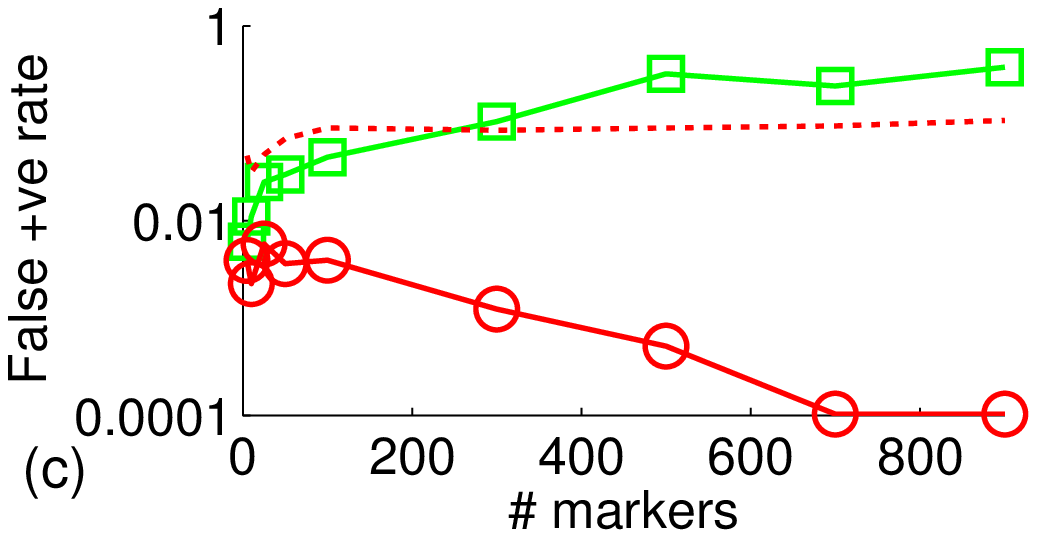} \label{fig:basic_evaluation_c}
  } 
  \caption{\textbf{Performance of Set Covering Reconstruction, relative to naive approaches and theoretical bounds.} For TPR, the data for naive 2 and set covering bound coincide. FPR for naive 2 is always zero, and hence not shown. Results are shown for networks of 100 nodes.}
  \label{fig:basic_evaluation}
\end{figure}

Figure \ref{fig:basic_evaluation} shows the results of network reconstruction using our Set Covering algorithm, along with baseline results and the worst-case bound. Probably the clearest result is that the bound on performance holds for both true and false positives, but more important is comparison of our algorithm with the naive baseline approaches.

Figures \ref{fig:basic_evaluation_b} and \ref{fig:basic_evaluation_c} clearly show that false positives are the cause of the poor performance of the first naive approach. This is entirely expected, but illustrates that it is not sufficient to simply find \textit{any} network that is consistent with the data. Both the first naive approach and our Set Covering algorithm return a network that is consistent with the data, but the results clearly show that searching for one that is maximally sparse leads to a reconstruction closer to the true network.

We also see in Fig.~\ref{fig:basic_evaluation} that the Set Covering algorithm exceeds the performance of the second naive approach. The second naive method never returns any false positives but throws away everything except the first two reports of every marker trace. This loses valuable information, and hence does not perform as well as the Set Covering algorithm.


\subsection{Jaccard Distance and MDL during set covering} \label{sec:jd_mdl_set_covering}

\begin{figure}
  \centering
  \subfloat{
  \includegraphics[width=0.5\textwidth]{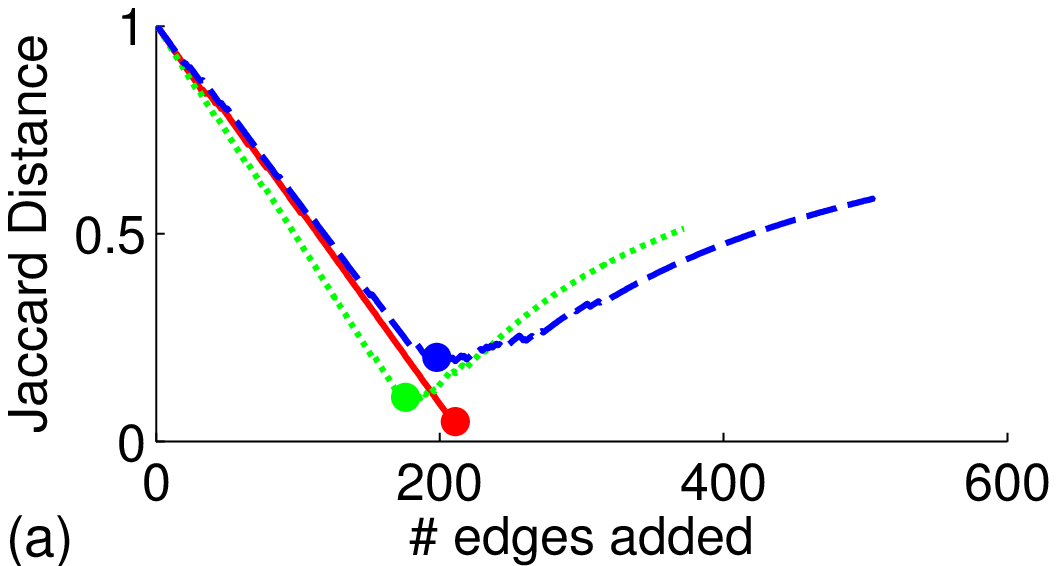} \label{fig:jd_mdl_vs_edges_a}
  }
  \subfloat{
  \includegraphics[width=0.5\textwidth]{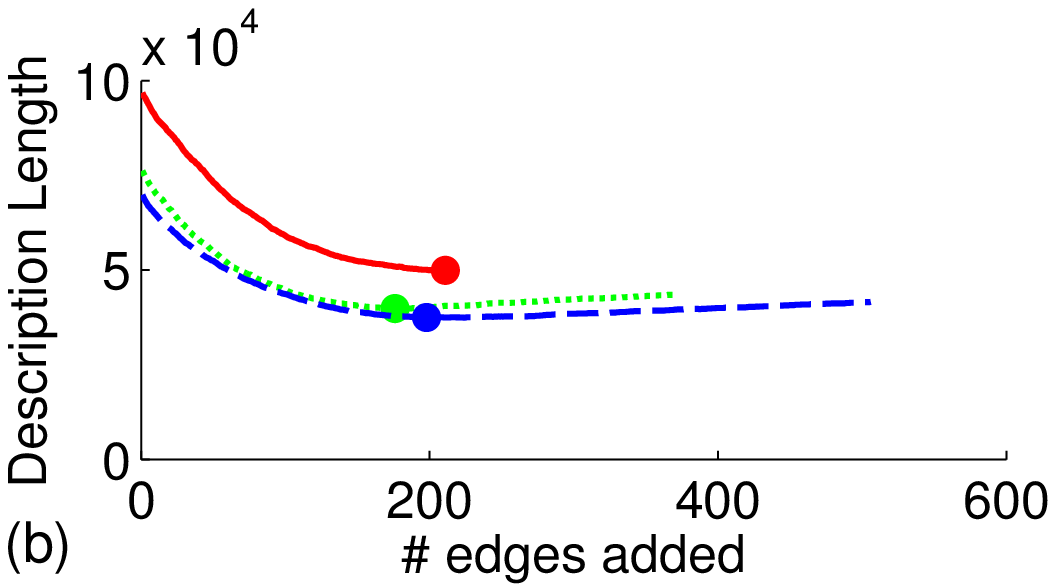} \label{fig:jd_mdl_vs_edges_b}
  }  \\
  \subfloat{
  \includegraphics[width=0.5\textwidth]{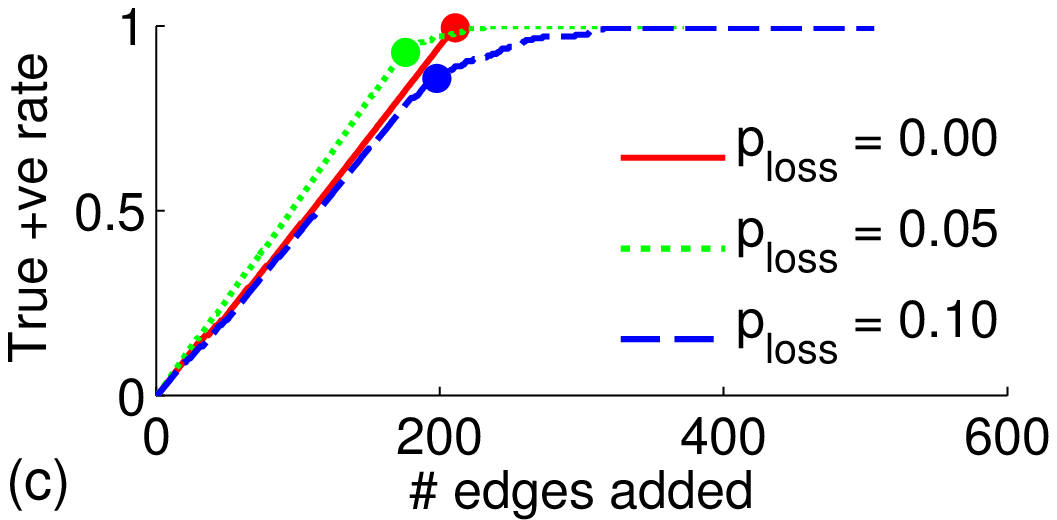} \label{fig:jd_mdl_vs_edges_c}
  } 
  \subfloat{
  \includegraphics[width=0.5\textwidth]{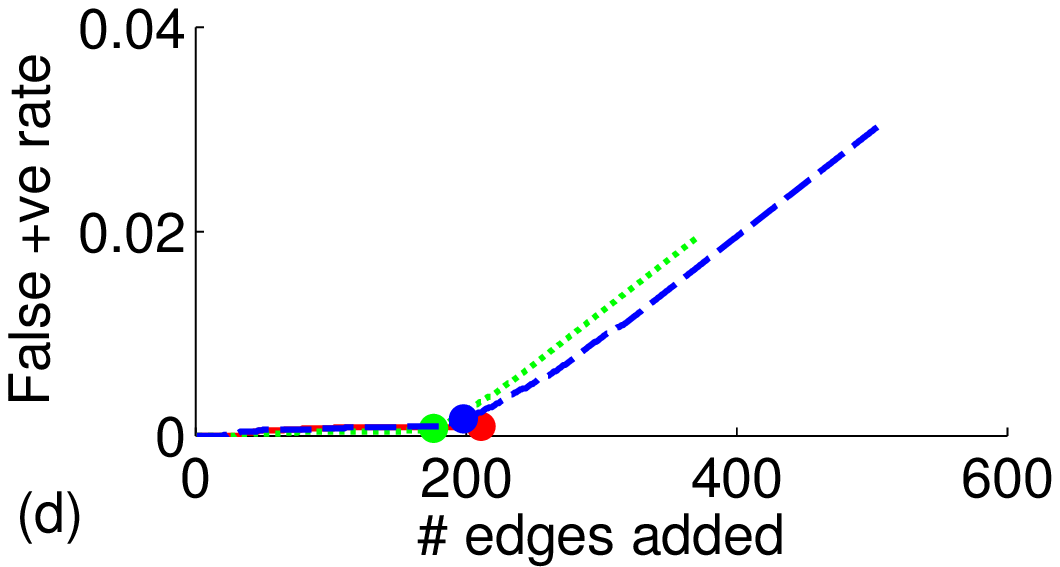} \label{fig:jd_mdl_vs_edges_d}
  } 
  \caption{\textbf{Plots showing variation of JD, DL, TPR and FPR with the progress of set covering.} Circles indicate the point at which the MDL criterion would have halted the covering. Each line shows reconstruction of a network of 100 nodes, using 1000 markers.}
  \label{fig:jd_mdl_vs_edges}
\end{figure}

In Sec.~\ref{sec:noisy_data} we introduced a criterion for early stopping, arguing that this would result in improved performance on noisy data. Figure \ref{fig:jd_mdl_vs_edges_a} gives the empirical verification for this, showing that for noisy data the closest match to the true network is obtained before the Set Covering is complete.

The circles plotted in Fig.~\ref{fig:jd_mdl_vs_edges} indicate the point at which the minimum description length was obtained, and hence the point at which the set covering would be halted. Figures \ref{fig:jd_mdl_vs_edges_c} and \ref{fig:jd_mdl_vs_edges_d} demonstrate that halting using MDL includes the majority of true positives, but limits the inclusion of false edges.

\subsection{Noisy Data with MDL Stopping}

\begin{figure}
  \centering
  \subfloat{
  \includegraphics[width=1.0\textwidth]{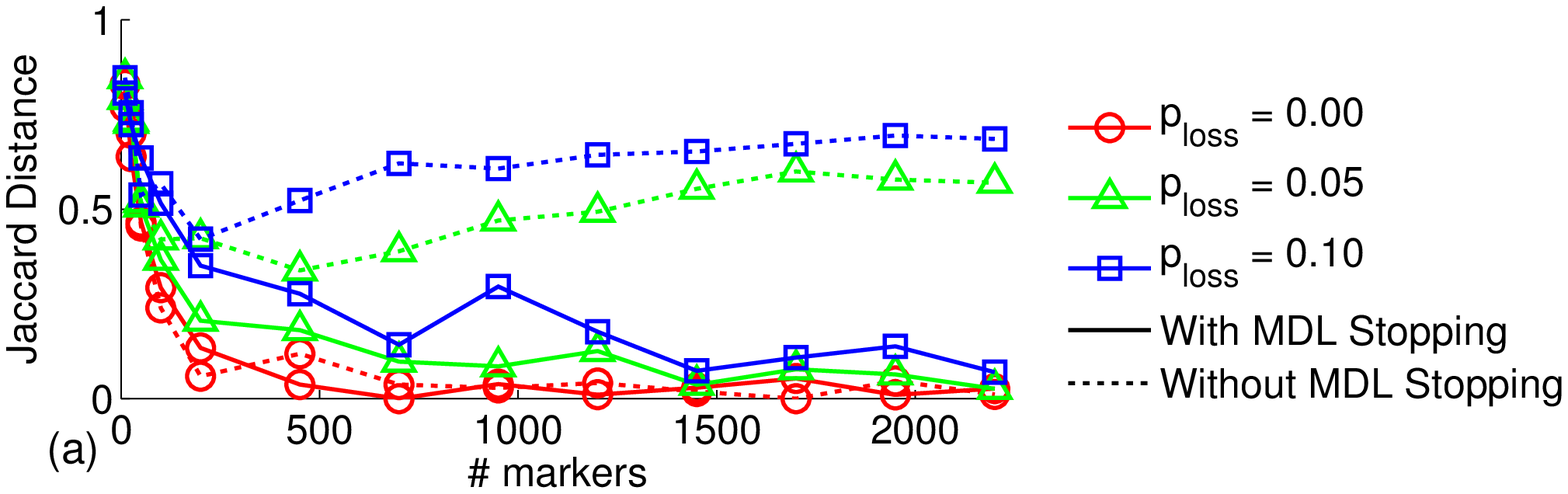} \label{fig:noisy_w_mdl_a}
  } \\
  \subfloat{
  \includegraphics[width=0.50\textwidth]{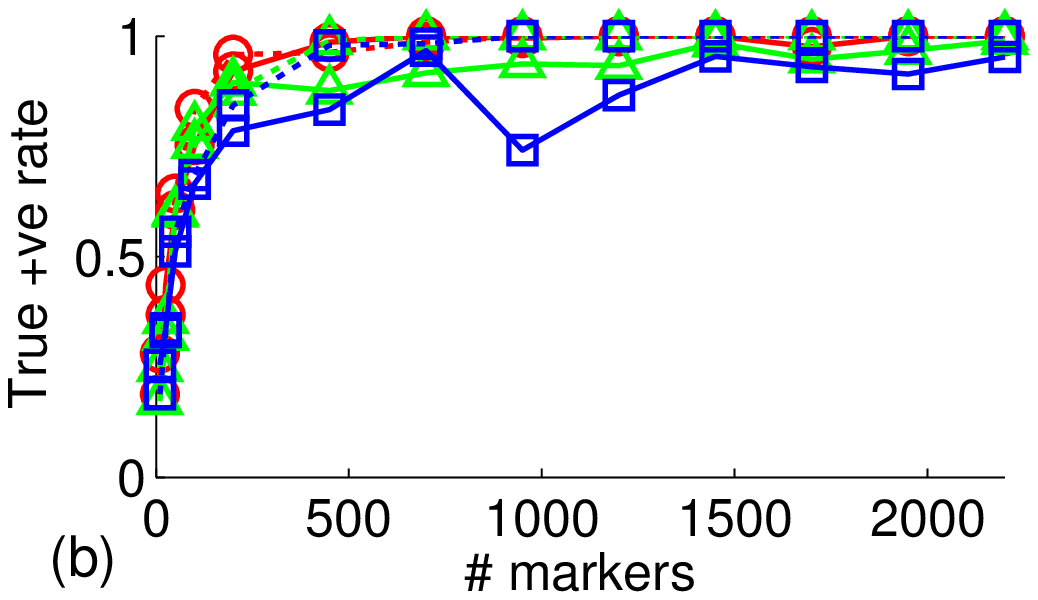} \label{fig:noisy_w_mdl_b}
  } 
  \subfloat{
  \includegraphics[width=0.50\textwidth]{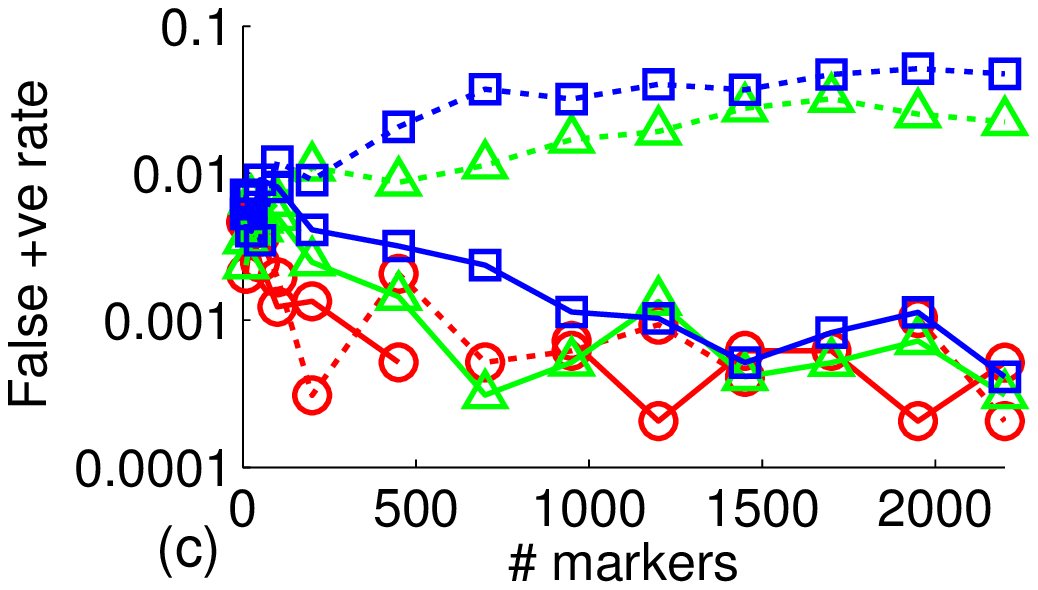} \label{fig:noisy_w_mdl_c}
  } 
  \caption{\textbf{Performance of reconstruction for various levels of noise, with and without MDL stopping.} Results are shown for a network of 100 nodes.}
  \label{fig:noisy_w_mdl}
\end{figure}

Finally, in Fig.~\ref{fig:noisy_w_mdl} we show results of network reconstruction for various noise levels, with and without the use of MDL stopping. Figure \ref{fig:noisy_w_mdl_c} clearly shows that when we use MDL stopping the rate of false positives remains bounded as the amount of data increases, in stark contrast to results for the basic algorithm.

The use of MDL does not completely compensate for the presence of noise, however, as evidenced by the lower performance shown in Fig.~\ref{fig:noisy_w_mdl_b}. The order in which edges are added to the set $E_R$ determines the proportion of true positives added before halting, and the results show that for higher noise conditions, more false edges will end up included in the final network. In the limit of large amounts of data, both TPR and FPR tend to a fixed level, determined by the exact nature of the network and markers, as well as the level of noise.

\section{Conclusions} \label{sec:conclusions}

Our work demonstrates a novel approach to the reconstruction of causal networks underlying stochastic branching processes, such as from data representing information flow or the spread of an epidemic on a network. Using the intuitive notion of consistency between a network and such data, we demonstrated that the entire network can be reconstructed node by node, using only local considerations. In this way, we were able to reformulate the problem in terms of the Set Covering problem, which is NP-hard but can be approximated well using an efficient greedy algorithm.

We developed two versions of the algorithm for different settings. The first version attempts to achieve perfect consistency with the data, and is therefore restricted to noise-free and fully-observed settings. This version is likely to be useful in controlled settings, such as in fault propagation networks in large enterprises. The second version was designed for the more common noisy setting, e.g. where certain marker observations may not have been observed or detected. It is based on the empirical observation that reliable edges tend to be added first, such that early stopping combined with our first algorithm is sufficient to provide good results. As a stopping criterion, the MDL principle proved to be an excellent measure, as shown by our experiments.

In further work we plan to investigate direct optimisation of the MDL cost function, rather than using MDL only as a stopping criterion. Another avenue for extending the approach is the use of exact times, rather than our current approach considering only the order of reports. Finally, we intend to apply our methods to various real-life data sets, such as the propagation of memes on the media network \cite{Leskovec:2009p5511,Snow1006:Finding}, and fault propagation data \cite{Rao:1987p6280}.

\bibliography{references_papers,references_additional}{}
\bibliographystyle{abbrv}

\subsubsection*{Acknowledgements.} 
The authors want to thank the `Pattern Analysis and Intelligent Systems' group at the University of Bristol for the benefit of numerous discussions. Nick Fyson is supported by the Bristol Centre for Complexity Sciences (EPSRC grant EP/5011214) and Nello Cristianini is supported by a Royal Society Wolfson Merit Award. This work is partially supported by EPSRC grant EP/G056447/1 and by the European Commission through the PASCAL2 Network of Excellence (FP7-216866).

%
%
%
%
%
%
%
%
%
%
%

\end{document}